\overfullrule=0pt
\input harvmac
\def\a{{\alpha}}
\def\ah{{\widehat\alpha}}
\def\ad{{\dot a}}
\def\bd{{\dot b}}
\def\l{{\lambda}}
\def\b{{\beta}}
\def\bh{{\widehat\beta}}
\def\g{{\gamma}}
\def\gh{{\widehat\gamma}}
\def\d{{\delta}}

\def\e{{\epsilon}}
\def\s{{\sigma}}
\def\N{{\nabla}}

\def\half{{1\over 2}}
\def\p{{\partial}}

\def\t{{\theta}}

\Title{\vbox{\hbox{IFT-P.072/2000 }}}
{\vbox{
\centerline{\bf Covariant Quantization of the Superstring}}}
\bigskip\centerline{Nathan Berkovits\foot{e-mail: nberkovi@ift.unesp.br}}
\bigskip
\centerline{\it Instituto de F\'\i sica Te\'orica, Universidade Estadual
Paulista}
\centerline{\it Rua Pamplona 145, 01405-900, S\~ao Paulo, SP, Brasil}

\vskip .3in
After reviewing the Green-Schwarz superstring using the approach
of Siegel, 
the superstring is covariantly quantized by constructing
a BRST operator from the fermionic constraints
and a bosonic pure spinor ghost variable. 
Physical massless vertex operators are constructed and, for
the first time, N-point tree
amplitudes are computed in a manifestly
ten-dimensional super-Poincar\'e covariant manner.
Quantization can be generalized to curved supergravity 
backgrounds and the vertex operator for fluctuations around 
$AdS_5\times S^5$ is explicitly constructed.
This review is written in a self-contained manner and is based
on talks given at the Fradkin Memorial Conference and Strings 2000.

\Date {August 2000}

\newsec{Introduction}

There are many motivations for covariantly quantizing the superstring.
As in any theory, it is desirable to make all
physical symmetries manifest in order to reduce the amount of
calculations and simplify any cancellations coming from the symmetry.
A second motivation comes from the desire to
construct a quantizable sigma model action for the superstring in 
curved backgrounds
with Ramond-Ramond flux. A third motivation is that a more symmetrical
formulation of superstring theory may shed light on some of the mysteries
of M-theory.

Although the light-cone Green-Schwarz (GS) formalism
\ref\GSlc{M.B. Green and J.H. Schwarz, {\it Supersymmetrical Dual
String Theory}, Nucl. Phys. B181 (1981) 502.}
has been useful for computing four-point tree-level and one-loop
amplitudes,
it is impractical for computing higher-point
or higher-loop amplitudes because of dependence on the locations of
the light-cone interaction points. Using the covariant RNS formalism,
one can easily compute N-point tree-level and one-loop amplitudes involving 
bosons, but amplitudes involving fermions are 
complicated by the presence of spin fields 
\ref\fms{D. Friedan, E. Martinec and S. Shenker,
{\it Conformal Invariance, Supersymmetry and String Theory},
Nucl. Phys. B271 (1986) 93.}. 
Furthermore, lack
of manifest spacetime supersymmetry in the RNS expressions 
complicates the analysis of finiteness properties and, for more than one-loop,
leads to picture-changing problems.

For more than 15 years, a classical 
super-Poincar\'e covariant
version of the GS formalism \ref\GS{M.B. Green and J.H. Schwarz,
{\it Covariant Description of Superstrings}, Phys. Lett. B136 (1984) 367.}
has existed. But until the recent work described here, 
quantization problems have prevented this
formalism from being used to compute non-vanishing scattering amplitudes.
In this review, the covariant GS formalism is quantized by constructing
a BRST operator from the fermionic constraints and
a bosonic pure spinor ghost variable.
After constructing physical massless vertex operators,
N-point tree
amplitudes are computed for the first time in a manifestly
ten-dimensional super-Poincar\'e covariant manner.

In previous papers, this author has discussed quantization of the 
superstring compactified to four \ref\four{ N. Berkovits, 
{\it Covariant Quantization Of
The Green-Schwarz Superstring in a Calabi-Yau Background},
Nucl. Phys. B431 (1994) 258, hep-th/9404162.} 
or six \ref\six{ N. Berkovits and C. Vafa,
{\it N=4 Topological Strings}, Nucl. Phys. B433 (1995) 123, hep-th/9407190.}
dimensions
using ``hybrid'' variables which combine 
four or six-dimensional GS variables with a $c=9$ or $c=6$
superconformal field theory for the compactification.
These hybrid formalisms have
manifest four or six-dimensional super-Poincar\'e covariance,
N=2 worldsheet supersymmetry, and are related to the usual RNS formalism
by a field redefinition. Unfortunately, the precise relation between
these hybrid formalisms and the new ten-dimensional super-Poincar\'e covariant
formalism is still unclear, so this review will not discuss the hybrid
formalisms.

In section 2 of this paper, the covariant GS formalism
will be reviewed using the approach of Siegel \ref\sieg{
W. Siegel, {\it Classical Superstring Mechanics}, Nucl. Phys. B263 (1986)
93.} where the canonical
momentum to $\theta^\a$ is an independent variable. 
In section 3, it will be argued \ref\cov{
N. Berkovits, {\it Super-Poincar\'e Covariant Quantization of the
Superstring}, JHEP 04 (2000) 018, hep-th/0001035.}
that a bosonic
pure spinor variable, $\l^\a$,
plays the role of the worldsheet ghost in this formalism, and a BRST
operator is constructed out of the fermionic constraints and $\l^\a$.
In section 4, physical massless vertex operators will be constructed
and, in section 5, they will be
used to compute N-point tree amplitudes \cov\ref\val{
N. Berkovits and B.C. Vallilo, {\it Consistency of Super-Poincar\'e
Covariant Superstring Tree Amplitudes}, JHEP 07 (2000) 015,
hep-th/0004171.}. These amplitudes are 
manifestly super-Poincar\'e covariant and involve
integration over an on-shell `harmonic' superspace including five $\t$'s and
three $\l$'s. In section 6, the cohomology of the BRST operator
is shown to reproduce the light-cone GS spectrum \ref\cohom{N. Berkovits,
{\it Cohomology in the Pure Spinor Formalism for the Superstring},
hep-th/0006003.}. In 
section 7, this quantization method is generalized to curved
supergravity backgrounds \cov\ and the vertex operator is constructed
for fluctuations around an $AdS_5\times S^5$ background with Ramond-Ramond
flux \ref\osv{N. Berkovits and Osvaldo Chand\'{\i}a, to appear.}. 
In section 8, some open problems and applications are discussed.

\newsec {Review of GS Formalism using the Approach of Siegel}

In conformal gauge, the classical
covariant GS action for the heterotic superstring is\GS
\eqn\het
{S=\int d^2 z [\half\Pi^m \bar\Pi_m +{1\over 4} 
\Pi_m \t^\a \g^m_{\a\b} \bar\p\t^\b
-{1\over 4}\bar\Pi_m \t^\a\g^m_{\a\b} \p\t^\b] + S_R}
where $x^m$ and $\t^\a$ are the $d=10$ worldsheet variables ($m=0$ to 9,
$\a=1$ to 16), $S_R$ describes the right-moving degrees of
freedom for the 
$E_8\times E_8$ or $SO(32)$ lattice, and
$\Pi^m = \p x^m + \half\t^\a \g^m_{\a\b} \p\t^\b$ and
$\bar\Pi^m = \bar\p x^m +\half \t^\a \g^m_{\a\b} \bar\p\t^\b$
are supersymmetric combinations of the momentum.
Note that
$\g^m_{\a\b}$ and $\g^{m\,\a\b}$ are $16\times 16$ symmetric matrices
satisfying 
$\g^{(m}_{\a\b} \g^{n)\,\b\g} = 2\eta^{mn}\d_\a^\g$ and are the
off-diagonal blocks in the Weyl representation of the 
$32\times 32$ ten-dimensional $\Gamma^m$-matrices.
In what follows, the right-moving degrees of freedom play no
role and will be ignored. Also, all of the following remarks
are easily generalized to the Type I and Type II superstrings.

Since the action of \het\ is in conformal gauge, it needs to be
supplemented with the Virasoro constraint $T= -\half \Pi^m \Pi_m=0$.
Also, since the canonical momentum to $\t^\a$ does not appear in the
action, one has the Dirac constraint 
$p_\a = {\d\L}/{\d \p_0\t^\a}=\half(\Pi_m -{1\over 4}\t\g_m\p_1\t) (\g^m\t)_\a$
where $p_\a$ is the canonical momentum to $\t^\a$. 
If one defines
\eqn\defdone{
d_\a = p_\a -\half(\Pi_m -{1\over 4}\t \g_m\p_1\t ) (\g^m\t)_\a,}
one can use the canonical commutation relations to find
$\{d_\a, d_\b\}= -\g^m_{\a\b} \Pi_m,$
which implies (since $\Pi^m\Pi_m=0$ is a constraint)
that the sixteen Dirac constraints $d_\a$ have eight
first-class components and eight second-class components.
Since the anti-commutator of the second-class constraints is
non-trivial (i.e. the
anti-commutator 
is an operator $\Pi^+$ rather than a constant), standard Dirac quantization
cannot be used since it would involve inverting an operator.
So except in light-cone gauge (where the commutator becomes a constant),
the covariant Green-Schwarz formalism cannot be easily quantized.

In 1986, Siegel suggested an alternative approach in which
the canonical momentum to $\t^\a$ is an independent variable
using the free-field action \sieg
\eqn\siegel
{S=\int d^2 z [\half \p x^m \bar\p x_m + p_\a\bar\p\t^\a].}
In this approach, Siegel attempted to replace 
the problematic constraints of the covariant GS action with some suitable
set of first-class constraints constructed out of the 
supersymmetric objects ($\Pi^m$, $d_\a$, $\p\t^\a$) where
\eqn\defdtwo{d_\a=p_\a - \half(\Pi^m -{1\over 4}\t\g^m\p\t) (\g_m\t)_\a}
is defined as in \defdone\ 
and is no longer constrained to vanish.
The first-class constraints should include the Virasoro constraint
$T=-\half \Pi^m \Pi_m
- d_\a \p\t^\a = -\half \p x^m \p x_m - p_\a \p\t^\a,$
and the first-class part of the $d_\a$ constraint, which is the 
irreducible part of 
$G^\a = \Pi^m (\g_m d)^\a.$ To get to light-cone gauge, one also needs
constraints such as $C^{mnp}= d_\a (\g^{mnp})^{\a\b} d_\b$ which is
supposed to replace the second-class constraints in $d_\a$. Although
this approach was successfully used for quantizing the superparticle
\ref\ilk{F. Essler, M. Hatsuda, T. Kimura, E. Laenen, 
A. Mikovic, W. Siegel and J.
Yamron, {\it Covariant Quantization of the First Ilk Superparticle},
Nucl. Phys. B364 (1991) 67.},
a set of constraints which closes at the quantum level and which
reproduces the correct physical superstring spectrum was never found.

Nevertheless, the approach of Siegel has the advantage that
all worldsheet fields are free which makes it trivial to compute
the OPE's that 
\eqn\opeone{x^m(y) x^n(z)\to -2\eta^{mn} \log|y-z|,\quad
p_\a(y)\t^\b(z) \to \d_\a^\b (y-z)^{-1}.}
 This gives some useful clues about the appropriate ghost
degrees of freedom. Since $(\t^\a,p_\a)$
contributes $-32$ to the conformal anomaly, the total
matter contribution is $-22$ which is expected to be cancelled by
a ghost contribution of $+22$.
Furthermore, the spin contribution to the $SO(9,1)$ Lorentz currents 
in Siegel's approach is $M_{mn}=\half p\g_{mn}\t$, as compared with
the spin contribution to the $SO(9,1)$ Lorentz currents
in the RNS formalism which is $\psi_m\psi_n$.
These two Lorentz currents satisfy similar OPE's except for the 
numerator in the double pole of $M_{mn}$ with $M_{mn}$, which
is $+4$ in Siegel's approach and $+1$ in the RNS formalism.
This suggests that the worldsheet ghosts should have Lorentz currents
which contribute $-3$ to the double pole.

\newsec{Quantization using Pure Spinors}

In fact, there exists an $SO(9,1)$ irreducible representation contributing
$c=22$ and with a $-3$ coefficient
in the double pole of its Lorentz current \cov.
This representation is a complex bosonic spinor $\l^\a$ satisfying the 
``pure spinor'' condition that 
\eqn\pure{\l^\a \g^m_{\a\b} \l^\b =0}
for $m=0$ to 9. Pure spinors have previously been used by Howe 
\ref\howe{P.S. Howe, {\it Pure Spinor Lines in Superspace and
Ten-Dimensional Supersymmetric Theories}, Phys. Lett. B258 (1991) 141,
Addendum-ibid.B259 (1991) 511\semi
P.S. Howe, {\it Pure Spinors, Function Superspaces and Supergravity
Theories in Ten Dimensions and Eleven Dimensions}, Phys. Lett. B273 (1991)
90.} 
for describing
super-Yang-Mills and supergravity equations of motion
as integrability conditions, and by
Nilsson \ref\Nil{B.E.W. Nilsson, ``Pure Spinors as Auxiliary Fields in
the Ten-Dimensional Supersymmetric Yang-Mills Theory'', Class. Quant. Grav.
3 (1986) L41.}
as superspace auxiliary fields. 

There are eleven independent complex degrees of freedom
in $\l^\a$, 
as can be seen by temporarily breaking $SO(9,1)$ to $SO(8)$ and solving
the constraint of \pure. If $\g^\pm =\half(\g^0\pm\g^9)$ and $\s^j_{a\ad}$
are the $SO(8)$ Pauli matrices satisfying $\s^{(j}_{a\ad} \s^{k)}_{a\bd}=
2 \d^{jk} \d_{\ad\bd}$ for $(j,a,\ad)=1$ to 8,
then 
$\l\g^-\l=0$ implies that $s^a =(\g^+\l)^a $ is a null complex $SO(8)$
spinor satisfying $s^a s^a=0$. 
Furthermore, $\l\g^j\l=0$ implies that 
$s^a \s^j_{a\ad}(\g^-\l)^\ad =0$, which implies that
$(\g^-\l)^\ad = v^j (\s^j s)^\ad$ for some complex vector $v^j$.
$\l\g^+\l=0$ gives no new constraints on $s^a$ and $v^j$.
But this parameterization of $\l^\a$ is invariant under
the gauge transformation $\d v^j = \e^\ad (\s^j s)^\ad$, which
allows one to gauge away half of the components of $v^j$. So a pure
spinor $\l^\a$ can be parameterized by the seven complex components
of a null $s^a$ and the four remaining complex components of $v^j$ \cohom.
After Wick rotation to $SO(10)$, one can alternatively describe
$\l^\a$ as a complex scale parameter multiplying the complex ten-dimensional
coset space $SO(10)/U(5)$ \cov.

Using
either the $SO(8)$ or $SO(10)/U(5)$ descriptions of $\l^\a$, 
a free field action $S_\l$ can be constructed out of the 
eleven left-moving
unconstrained parameters and their canonical momenta with 
conformal anomaly $c=22$. 
Although the unconstrained parameters
do not transform covariantly under $SO(9,1)$, the only combinations which 
appear in the vertex operators are $\l^\a$ and its Lorentz current
$N_{mn}$, which satisfy the manifestly
SO(9,1) covariant OPE's
\eqn\opetwo{N_{mn}(y)\l^\a(z) \to { \half{(\g^{mn}\l)^\a}\over{y-z}},
\quad N_{kl}(y) N_{mn}(z) \to
{{\eta_{m[l} N_{k]n} - 
\eta_{n[l} N_{k]m} }\over {y-z}} - 3
{{\eta_{k[n} \eta_{m]l} }
\over{(y-z)^2}}  .}
One can alternatively obtain the
OPE's of \opetwo\ from an SO(9,1) WZW model of level $-3$, however, $\l^a$
would not be a fundamental field in such an action \ref\bersh
{M. Bershadsky, private communication.}.

One still needs to define which states
are physical in this Hilbert space. The physical state condition
will be defined as ghost-number one states in the cohomology of
the BRST-like operator \cov 
\eqn\defq{Q= \int dz \l^\a(z) d_\a(z)}
where $\l^\a$ carries ghost-number one and $d_\a$ is defined in \defdtwo.
Since $d_\a$ satisfies the OPE $d_\a(y) d_\b(z) \to -(y-z)^{-1} \g^m_{\a\b}
\Pi_m$, the pure spinor condition on $\l^\a$ implies that $Q$ is nilpotent.
Although it is a bit unusual that $Q$ is constructed from second-class
constraints, it will be shown in section 5 that its cohomology reproduces
the correct light-cone GS spectrum. 

\newsec{ Physical Massless Vertex Operators}

Since there is no tachyon, the massless open superstring
state is constructed from
zero modes of the dimension-zero
worldsheet fields $x^m$,$\t^\a$, and $\l^\a$.
Since it has ghost-number one, it can be written as \cov
\eqn\unint{U= \l^\a A_\a (x,\t)}
where $A_\a(x,\t)$ is a spinor superfield.
Since $d_\a(y) A_\b(x,\t) \to (y-z)^{-1} D_\a A_\b(x,\t)$
where $D_\a = {\p\over{\p\t^\a}} +(\g^m\t)_\a \p_m$ is the
superspace derivative, $QU=0$ implies that
$\l^\a\l^\b D_\a A_\b =0$. But since $\l^\a\l^\b = {1\over{32}}
(\g^{mnpqr})^{\a\b} (\l\g_{mnpqr}\l),$ this implies that 
$(\g_{mnpqr})^{\a\b} D_\a A_\b =0$ for every five-form direction $mnpqr$.
Furthermore, the gauge invariance $\d U = Q\Omega$ implies that
$A_\a$ has the gauge transformation $\d A_\a = D_\a \Omega$.
 
The equation 
$(\g_{mnpqr})^{\a\b} D_\a A_\b =0$ and gauge invariance
$\d A_\a = D_\a \Omega$ describe the spinor gauge superfield
for linearized on-shell
$d=10$ super-Yang-Mills \ref\sym
{W. Siegel, {\it Superfields in Higher-Dimensional Spacetime},
Phys. Lett. 80B (1979) 220.}.
It can be gauge-fixed to the form
\eqn\gf{
A_\a(x,\t) = a_m(x) (\g^m\t)_\a + (\xi(x) \g_{mnp})_\a (\t\g^{mnp}\t) + ...}
where $a_m(x)$ and $\xi^\a(x)$ are the on-shell gluon and gluino and
all the components in 
$...$ are auxiliary fields which are related to $a_m$ and $\xi^\a$ by
equations of motion. 

To compute scattering amplitudes, one also needs vertex operators
in integrated form, $\int dz V$, where $V$ is 
usually obtained from the unintegrated
vertex operator $U$ by anti-commuting with the $b$ ghost. But since
there is no natural candidate for the $b$ ghost in
this formalism, 
one needs to use an alternative method for obtaining $V$ which is
from the relation
$[Q,V] = \p U$ \ref\bp
{N. Berkovits, M. Hatsuda and W. Siegel, {\it
The Big Picture}, Nucl. Phys. B371 (1992) 434, hep-th/9108021.}.
Using this alternative method, one finds for the open
superstring massless vertex operator that \cov 
\eqn\intv{
V= A_\a(x,\t) \p\t^\a + A_m(x,\t) \Pi^m + W^\a(x,\t) d_\a + F^{mn}(x,\t)
N_{mn},}
where $A_m = {1\over{8}}\g_m^{\a\b} D_\a A_\b$, 
$W^\a = {1\over {10}} \g_m^{\a\b} (D_\b A^m - \p^m A_\b)$,
and 
$F^{mn}= \p^{[m} A^{n]}$.
It will be useful to note that in components,
\eqn\rns{
V = a_m (x) \p x^m + \p_{[m} a_{n]}(x) M^{mn} + \xi^\a(x) q_\a + O(\t^2),}
where $M^{mn}=
\half p\g^{mn}\t + N^{mn}$ is the spin contribution to the Lorentz current
and
$q_\a= p_\a +\half(\p x^m +{1\over{12}}\t\g^m\t) (\g_m\t)_\a$
is the spacetime-supersymmetry current,
so \rns\ closely resembles the RNS vertex operator \fms\ for the gluon
and gluino. If one drops the $F^{mn} N_{mn}$ term, the vertex operator
of \intv\ was suggested 
by Siegel \sieg\ based on superspace arguments. 

For the closed superstring, the massless vertex operator is
$U=\l^\a \widehat\l^{\bh} A_{\a\bh}(x,\t,\widehat\t)$ where
$\widehat\l^\ah$ and $\widehat\t^{\ah}$ are right-moving worldsheet fields
and the chirality of the
$\widehat\a$ index depends if the superstring is IIA or IIB.
The physical state condition $QU=\widehat QU =0$ and gauge invariance
$\d U = Q\widehat\Omega + \widehat Q\Omega$ 
where $\widehat Q\widehat\Omega=Q\Omega=0$
implies that \osv
\eqn\eom{\g_{mnpqr}^{\a\b} D_\a A_{\b\gh} =  
\g_{mnpqr}^{\ah\gh} \widehat D_\ah A_{\b\gh} = 
0, \quad \d A_{\a\bh} = D_\a \widehat\Omega_\bh + \widehat D_\bh \Omega_\a,}
$$\g_{mnpqr}^{\a\b} D_\a \Omega_\b =  
\g_{mnpqr}^{\ah\gh}\widehat D_\ah \widehat\Omega_\gh = 0$$
for any five-form direction
$mnpqr$, which are the linearized equations of motion
and gauge invariances 
of the Type IIA or Type IIB supergravity multiplet. The integrated form
of the closed superstring massless vertex operator 
is the left-right product of
the open superstring vertex operator of \intv\ and will be used in section 7
for quantizing the superstring in a curved supergravity background. 

\newsec { Tree-Level Massless Scattering Amplitudes}

As usual, the $N$-point tree-level open superstring
scattering amplitude will be defined as
the correlation function of $3$ unintegrated vertex operators $U_r$
and $N-3$ integrated vertex operators $\int dz V_r$.
For massless external states, these vertex operators are given in \unint\
and \intv. 

The first step to evaluate the correlation function
is to eliminate all worldsheet fields of non-zero dimension
(i.e. $\p x^m$, $\p\t^\a$, $p_\a$ and $N^{mn}$)
by using their OPE's with other worldsheet fields and 
the fact that they vanish at $z\to\infty$.
One then integrates over the $x^m$ zero modes
to get a Koba-Nielson type formula, 
\eqn\koba{A = \int dz_4 ... dz_N  \langle \l^\a \l^\b \l^\g
f_{\a\b\g}(z_r, k_r, \eta_r, \t) \rangle}
where $\l^\a\l^\b\l^\g$ comes from the three unintegrated
vertex operators and $f_{\a\b\g}$ is some function of
the $z_r$'s, the momenta $k_r$, the polarizations 
$\eta_r$, and the remaining $\t$ zero modes.

One would like to define the correlation function 
$\langle \l^\a \l^\b \l^\g f_{\a\b\g}\rangle$
such
that $A$ is supersymmetric and gauge invariant. An obvious way to
make $A$ supersymmetric is to require that the correlation function
vanishes unless all sixteen $\t$ zero modes are present, but this
gives the wrong answer by dimensional analysis. The correct answer
comes from realizing that $Y= \l^\a\l^\b\l^\g f_{\a\b\g}$
satisfies the constraint $QY=0$ since the external states are on-shell.
Furthermore, gauge invariance implies that $\langle Y\rangle$ should
vanish whenever $Y=Q\Omega$. 

At zero momentum and ghost-number three, there is precisely one
state in the cohomology of $Q$ which is $(\l\g^m\t)(\l\g^n\t)(\l\g^p\t)
(\t\g_{mnp}\t)$. So if 
\eqn\expa{f_{\a\b\g}(\t) = A_{\a\b\g}+ \t^\d B_{\a\b\g\d} + ... +
(\g^m\t)_\a (\g^n\t)_\b (\g^p\t)_\g (\t \g_{mnp}\t) F + ...,}
it is natural to define 
\eqn\harm{\langle \l^\a \l^\b \l^\g f_{\a\b\g}(z_r,k_r,\eta_r,\t)\rangle
 = F(z_r,k_r,\eta_r) .}
This definition is supersymmetric since 
$(\l\g^m\t)(\l\g^n\t)(\l\g^p\t)(\t \g_{mnp}\t) \t^\a$ is not annihilated
by $Q$, and is gauge invariant since 
$(\l\g^m\t)(\l\g^n\t)(\l\g^p\t)(\t \g_{mnp}\t)\neq Q\Omega$. Note that
\harm\ can be interpreted as integration over an on-shell harmonic
superspace involving five $\t$'s since
$\langle 
\l^\a\l^\b\l^\g f_{\a\b\g}\rangle = \int (d\t \g_m)^\a (d\t \g_n)^\b
(d\t \g_p)^\g (d\t\g^{mnp}d\t) f_{\a\b\g}$ \cov.

For three-point scattering, $A=\langle \l^a A_\a^1 \l^\b A_\b^2 \l^\g A_\g^3
\rangle$, it is easy to check that the prescription of \harm\ reproduces
the usual super-Yang-Mills cubic vertex. In the gauge of \gf, each $A_\a$
contributes one, two or three $\t$'s. If the five $\t$'s are distributed
as $(1,1,3)$, one gets the $a_m^1 a_n^2 \p^{[m} a^{3 n]}$ vertex,
whereas if they are distributed as $(2,2,1)$, one gets the 
$(\xi^1 \g^m\xi^2) a_m^3$ vertex. Together with Brenno Vallilo,
it was proven that the above prescription agrees with the standard RNS
prescription of \fms\
for N-point massless tree amplitudes involving up to four
fermions \val. The relation of \rns\ to the RNS massless vertex operator was
used in this proof, and the restriction on the number of fermions comes
from the need for different pictures in the RNS prescription.

\newsec{ Cohomology of Q}

To compute the cohomology of $Q=\int dz \l^\a d_\a$, it is convenient
to use the $SO(8)$ parameterization of $\l^\a$ in terms of $s^a$ and
$v^j$ discussed in section 3. The gauge invariance $\d v^j = \e^\ad
(\s^j s)^\ad$ leads to a new fermionic ghost anti-chiral
spinor $t^\ad$, and the
invariance of the gauge parameter, $\d\e^\ad = y^j (\s^j s)^\ad$,
leads to a new bosonic ghost-for-ghost vector $v_{(1)}^j$. 
These gauge invariances continue indefinitely until one has an
infinite chain of bosonic vectors $v^j_{(n)}$ and 
fermionic anti-chiral spinors $t^j_{(n)}$ for $n=0$ to $\infty$
where $n=0$ labels the original vector and spinor \cohom.

In terms of these left-moving dimension-zero worldsheet fields, 
$Q = \int dz  s^a G^a$ where
$$G^a = (\g^- d)^a + \s_j^{a\ad}[ v^j_{(0)} (\g^+ d)^\ad +
\sum_{n=0}^\infty (w_{(n)}^j t^\ad_{(n)} + v^j_{(n+1)} u^\ad_{(n)})],$$
$(w_{(n)}^j, u^\ad_{(n)})$ are the canonical momenta
for $(v_{(n)}^j, t^\ad_{(n)})$, and the infinite sum comes from the
gauge invariances. Note that $Q^2=0$ since
$s^a s^a =0$ and
$G^a(y) G^b(z) \to 2 (y-z)^{-1}\d^{ab} T(z)$ where
$$T =\half\Pi^- + v^j\Pi^j +\half
v^j v^j \Pi^+ + t^\ad_{(0)} (\g^+ d)^\ad
+\sum_{n=0}^\infty (v^j_{(n+1)} w^j_{(n)} + t^\ad_{(n+1)} u^\ad_{(n)}).$$

One can treat $s^a s^a=0$ as a BRST constraint by defining
$Q' = \int dz [ s^a G^a - b s^a s^a + c T ] $
where $(b,c)$ are fermionic ghosts of dimension $(1,0)$.
One can check that $Q'$ is nilpotent with unconstrained $s^a$, and has
cohomology equal to that of $Q$.  Note that
the algebra of $G^a$ and $T$ is not the usual $N=8$ superconformal algebra
since, for example, $T$ has dimension 1 and commutes with $G^a$.
Nevertheless, since $T$ and $G^a$ are first-class constraints, they
can be used to gauge-fix $x^+=P^+\tau$ and $(\g^+\t)^a=0$, and solve for
$x^-$ and $(\g^- p)^a$. 

The only remaining constraint is that physical operators must commute with
the zero mode of $T$, which in this light-cone gauge is $T_0 + \half P^-$
where
$$T_0 = \int dz [-\half \t\g^-\p\t + v^j_{(0)}\p x^j + \half v^j_{(0)}
v^j_{(0)} P^+ + t_{(0)}^\ad (\g^+ d)^\ad +\sum_{n=0}^\infty
(v^j_{(n+1)} w^j_{(n)} + t^\ad_{(n+1)} u^\ad_{(n)})]$$
and a Lorentz frame has been chosen in which $P^j=0$ for $j=1$ to 8.
So physical operators are constructed from products of eigenvectors of $T_0$
whose $(mass)^2 = P^+ P^-$ is equal to the sum of the eigenvalues
multiplied by $-2 P^+$.
Since $T_0$ is quadratic, the bosonic and fermionic
eigenvectors with eigenvalues $N$, $a_N$ and $b_N$, can be expressed as linear
combinations of the remaining light-cone variables,
$$a_N = \int d\sigma [f_N^j \p x^j + \sum_{n=0}^\infty (g_{N(n)}^j v_{(n)}^j
+ h^j_{N(n)} w^j_{(n)})],$$
$$b_N = \int d\sigma [j_N^\ad (\g^+ p)^\ad + k_N^\ad (\g^-\t)^\ad
+ \sum_{n=0}^\infty (l_{N(n)}^\ad t_{(n)}^\ad
+ m^\ad_{N(n)} u^\ad_{(n)})].$$
If one requires that these eigenvectors satisfy the normalization condition
that $\int d\s (f_N^j f_N^j + \sum_{n=0}^\infty g_{N(n)}^j h_{N(n)}^j)$
and 
$\int d\s (j_N^\ad k_N^\ad + \sum_{n=0}^\infty l_{N(n)}^\ad m_{N(n)}^\ad)$
are finite, one finds that the only normalizable eigenvectors are
the modes of
$$y^j = \p x^j + \sum_{n=0}^\infty (P^+)^{-n-1} \p^{n+1} w_{(n)}^j,\quad
q^\ad = (\g^+ p -\half P^+\g^-\t)^\ad +\sum_{n=0}^\infty (P^+)^{-n-1} 
\p^{n+1} u_{(n)}^\ad.$$ 
These eight bosonic and eight fermionic eigenvectors generate the
usual light-cone GS spectrum. So the $T_0$ constraint
has imposed the
second-class constraints as well as the mass-shell condition.
It is interesting to note that an infinite set of fields has
also been useful for treating the second-class constraints of the
chiral boson \ref\mwy{
B. McClain, Y.S. Wu and F. Yu,
{\it Covariant Quantization of Chiral Bosons and OSp(1,1/2) Symmetry},
Nucl. Phys. B343 (1990) 689\semi C. Wotzasek, {\it The Wess-Zumino Term
for Chiral Bosons}, Phys. Rev. Lett. 66 (1991) 129.} 
and the self-dual Type IIB four-form \ref\fourform{
N. Berkovits,
{\it Manifest Electromagnetic Duality in Closed
Superstring Field Theory}, Phys. Lett. B388 (1996) 743, hep-th/9607070.}.

\newsec{ Superstring in $AdS_5\times S^5$ Background}

To construct the superstring action in a curved Type II supergravity
background, one adds the integrated form of
the closed massless vertex operator of section 4
to the superstring
action in a flat background, and covariantizes with respect
to super-reparameterization invariance. As usual, one also
needs to include a Fradkin-Tseytlin term for coupling the dilaton
to the worldsheet curvature $r$. The resulting action is \cov\ref\eff
{N. Berkovits and W. Siegel, {\it
Superspace Effective Actions for 4D Compactifications
of Heterotic and Type II Superstrings},
Nucl. Phys. B462 (1996) 213, hep-th/9510106.}
\eqn\curv{S=\int d^2 z [\half
\p Y^M \bar\p Y^N 
(G_{MN}(Y)+ B_{MN}(Y)) }
$$ 
+
\bar\p Y^M \widetilde d_\a  E^\a_M(Y)
+\p Y^M \widetilde{\widehat d}_{\widehat\a} E^{\widehat\a}_M (Y)
+ \widetilde d_{\a} \widetilde{\widehat d}_{\widehat\b}
F^{\a \widehat\b }(Y) + \a' r \Phi(Y)] + S_\l + S_{\widehat\l} $$
where $Y^M = (x^m,\t^\mu,\widehat\t^{\widehat\mu})$ 
parameterizes the curved N=2
superspace background, the first line of \curv\ is the usual GS action, 
$\widetilde d_\a = d_\a + N_{mn} (\g^{mn} D)_\a$ and
$\widetilde {\widehat d}_\ah = \widehat d_\ah + \widehat N_{mn} (\g^{mn} 
\widehat D)_\ah$ where $D_\a$ and $\widehat D_\ah$ are $N=2$ superspace
derivatives which act on the background fields to their right,
$S_\l$ and $S_{\widehat\l}$ are the actions in a flat background for $\l^\a$
and $\widehat\l^\ah$, and
the lowest components of the superfields 
$E_m^\a$ and $E_m^{\widehat\a}$ are the gravitini, 
of $F^{\a\bh}$ are the Ramond-Ramond field strengths, and of $\Phi$ is
the dilaton. It is convenient to
treat $d_\a$ and $\widehat d_\ah$ (instead of $p_\a$
and $\widehat p_\ah$) 
as fundamental worldsheet
fields when the background is curved.
Although the above action does not have $\kappa$-symmetry, it
is constructed such that $\l^\a d_\a$ and $\widehat\l^\ah \widehat d_\ah$
are holomorphic and anti-holomorphic currents when the background
superfields are on-shell. One can therefore
quantize the action using $Q=\int dz \l^\a d_\a$ and
$\widehat Q=\int d\bar z \widehat \l^\ah \widehat d_\ah$ as in the flat case.

In the $AdS_5 \times S^5$ background with $n$ units of Ramond-Ramond flux,
$F^{\a\bh}= n\g_{01234}^{\a\bh} = n\d^{\a\bh}$, so the term
$n\d^{\a\bh} d_\a \widehat d_\bh$ in \curv\ allows one to solve
the equations of motion \ref\witten{ N. Berkovits, C. Vafa
and E. Witten, {\it Conformal Field Theory of AdS Background
with Ramond-Ramond Flux}, JHEP 9903 (1999) 018, hep-th/9902098.}
for $d_\a$ and $\widehat d_\ah$ in terms of
$(x^m,\t^\a,\widehat\t^\ah)$. Plugging in the appropriate values
for the $AdS_5\times S^5$ background superfields, one obtains \cov\ref\bz
{N. Berkovits, M. Bershadsky, 
T. Hauer, S. Zhukov and B. Zwiebach, {\it Superstring Theory on
$AdS_2\times S^2$ as a Coset Supermanifold}, Nucl. Phys. B567 (2000) 61, 
hep-th/9907200.}
\ref\Metsaev{R. Metsaev and A. Tseytlin,
{\it Type IIB superstring action in $AdS_5 \times S^5$
background},
Nucl.Phys. {B533} (1998) 109, hep-th/9805028.}
$$S= {1\over{n^2 g_s^2}}
\int d^2 z [ \half\eta_{cd} J^c \bar J^d +{1\over 4} \d_{\a\bh} 
(3 J^\bh \bar J^\a - J^\a \bar J^\bh) + N_{cd} J^{cd} +\widehat N_{cd}
\bar J^{cd} ] + S_\l + S_{\widehat\l}$$
where $J^A = (g^{-1} \p g)^A$ and $\bar J^A = (g^{-1} \bar\p g)^A$ with
$A=(a,\a,\ah,cd)$ are
left-invariant currents constructed from the coset supergroup 
$g(x,\t,\widehat\t) \in {{PSU(2,2|4)}\over{SO(4,1)\times SO(5)}}$.
This action is invariant under $g\to M g \Omega$ where $M$ is
a global $PSU(2,2|4)$ transformation and $\Omega$ is a local 
$SO(4,1)\times SO(5)$ transformation which also acts as
a Lorentz rotation on the 
pure spinors $\l^\a$ and $\widehat\l^\ah$.
It has been checked to one-loop order \bz\ \bersh\ 
that $S$ is conformally invariant
and that $\l^\a d_\a = \d_{\a\bh}\l^\a J^\bh$
and $\widehat\l^\ah \widehat d_\ah = \d_{\a\bh}\widehat\l^\bh\bar J^\a$
are holomorphic and anti-holomorphic currents.

To construct the vertex operator $U=\l^\a\widehat\l^\bh A_{\a\bh}$
for fluctuations around the $AdS_5\times S^5$ background, one needs
to generalize the equations of motion and gauge invariances of \eom\
as was done in 
\ref\dw{L. Dolan and E. Witten, {\it Vertex Operators
for $AdS_3\times S^3$ Background with Ramond-Ramond Flux}, JHEP 9911
(1999) 003, hep-th/9910205.} for the $AdS_3\times S^3$ background.
The flat space equations of \eom\ can be generalized to $AdS_5\times
S^5$ by simply replacing
$D_\a$ and $\widehat D_\ah$ in \eom\ with $\N_\a = E_\a^M (\p_M + \omega_M)$
and $\widehat\N_\ah = E_\ah^M (\p_M + \omega_M)$ where 
$\omega_M^{cd} = (g^{-1} \p_M g)^{cd}$ is the $SO(4,1)\times SO(5)$
spin connection and $E_A^M$ is the super-vierbein obtained by
inverting $E_M^A = (g^{-1} \p_M g)^A$ for $A= (a,\a,\ah)$.
The equations of motion and gauge invariances are still self-consistent
since, although $\{\N_\a,\widehat\N_\bh\}$ is non-zero in the $AdS_5\times S^5$
background, one can check that
$\g_{mnpqr}^{\a\g}
\{\N_\a,\widehat\N_\bh\} f_\g = 0$ and 
$\g_{mnpqr}^{\bh\gh}
\{\N_\a,\widehat\N_\bh\} \widehat f_\gh = 0$
for any $f_\g$ and $\widehat f_\gh$ and 
for any five-form direction $mnpqr$ \osv.

\newsec{Open Problems and Applications}

In this review, it was argued that pure spinors are the
worldsheet ghosts of the GS formalism and that physical states are 
described by ghost-number one vertex operators
in the cohomology of $Q=\int dz \l^\a d_\a$. Although
these ideas were successfully used for computing N-point tree amplitudes, 
a more
geometrical understanding would be useful for computing loop amplitudes
where the ghosts play a more important role. 
In particular, one would like to understand how to formulate the
action in a reparameterization-invariant manner and how the ghosts
arise from the gauge-fixing procedure.

The relatively simple form of the vertex operator for fluctuations
around the $AdS_5\times S^5$ background suggests that it might
be possible to compute tree amplitudes in this background. Although
the currents $J^A$ are not holomorphic since $\bar\p J^A = 
h^A_{BC} J^B \bar J^C$ where $h^A_{BC}$ are constants \ref\btwo{M. Bershadsky,
S. Zhukov and A. Zaintrob, {\it $PSL(n|n)$ Sigma Model as a Conformal
Field Theory}, Nucl. Phys. B559 (1999) 205, hep-th/9902180.} \witten,
conformal invariance together with the $AdS$ isometries may be enough
to imply their OPE's.
Of course, even knowing these OPE's,
one would still have to compute the zero mode contribution
from $(x^m,\t^\mu,\widehat\t^{\widehat\mu})$ to the tree amplitude,
which might be complicated in an $AdS_5\times S^5$ background \ref\wpc
{E. Witten, private communication.}.

{\bf Acknowledgements:}
I would like to thank Michael Bershadsky, Osvaldo Chand\'{\i}a,
Warren Siegel, Brenno Vallilo, and Edward Witten for useful
conversations, and
CNPq grant 300256/94-9 and the Clay Mathematics Institute
for partial financial support. 
This research was partially conducted during the period the author
was employed by the Clay Mathematics Institute as a CMI Prize Fellow.

\listrefs

\end